**Online information of vaccines: information quality is an ethical responsibility of search engines**

Pietro Ghezzi[1*], Peter G Bannister[1], Gonzalo Casino[2-3], Alessia Catalani[4], Michel Goldman[5], Jessica Morley[6], Marie Neunez[5], Andreu Prados[2-7], Mariarosaria Taddeo[6-8], Tania Vanzolini[4] and Luciano Floridi [6-8]

[1] Brighton & Sussex Medical School, Falmer, Brighton, UK

[2] Communication Department, Pompeu Fabra University, Barcelona, Spain

[3] Iberoamerican Cochrane Center, Barcelona, Spain

[4] Department of Biomolecular Sciences, University of Urbino Carlo Bo, Urbino, PU, Italy.

[5] Institute for Interdisciplinary Innovation in healthcare (I3h), Université libre de Bruxelles

[6] Oxford Internet Institute, University of Oxford, Oxford, UK

[7] Blanquerna School of Health Sciences, Universitat Ramon Llull, Barcelona, Spain.

[8] The Alan Turing Institute, London, UK

*Corresponding author: Professor Pietro Ghezzi, Brighton and Sussex Medical School, Trafford Centre, University of Sussex, Falmer, Brighton, BN19RY, UK. Email: p.ghezzi@bsms.ac.uk; phone: +44-(0)1273-873112

**Abstract**

The fact that Internet companies may record our personal data and track our online behaviour for commercial or political purpose has emphasized aspects related to online privacy. This has also led to the development of search engines that promise no tracking and privacy. Search engines also have a major role in spreading low-quality health information such as that of anti-vaccine websites.

This study investigates the relationship between search engines' approach to privacy and the scientific quality of the information they return. We analyzed the first 30 webpages returned searching "vaccines autism" in English, Spanish, Italian and French. The results show that "alternative" search engines (Duckduckgo, Ecosia, Qwant, Swisscows and Mojeek) may return more anti-vaccine pages (10-53%) than Google.com (0%). Some localized versions of Google, however, returned more anti-vaccine webpages (up to 10%) than Google.com.

Our study suggests that designing a search engine that is privacy savvy and avoids issues with filter bubbles that can result from user-tracking is necessary but insufficient; instead mechanisms should be developed to test search engines from the perspective of information quality (particularly for health-related webpages), before they can be deemed trustworthy providers of public health information.

**Introduction**

The World Health Organization lists vaccine hesitancy as one of the top ten threats to global health in 2019 [1], requiring ongoing global monitoring. Despite the fact that the 1998 study which incorrectly suggested that the MMR vaccine could cause autism in children and prompted antivaccine beliefs [2] has now been discredited, misinformation, and indeed, disinformation[1], about vaccines continues to be published on the Internet perpetuating such beliefs.

It has been suggested that this misinformation plays a role in the current low uptake of vaccines in developed countries [3]. Understanding whether this is actually the case, and how to address this issue, is crucial as web-based sources of health information may be instrumental to solve the sustainability challenge currently facing health systems across the globe [4].

---

[1] Misinformation is incorrect information that stems from human error e.g. a lack of fact checking, whereas disinformation is purposefully and deliberately incorrect. Both matter in this context because whilst the original 1998 study may have been an example of misinformation, it is possible that maleficent actors are now purposefully spreading disinformation for the purpose of undermining public health.

The accuracy of information provided by a website is a key indicator of its overall information quality (IQ). The broader aspects of IQ have been the subject of many studies [5], but health IQ and trustworthiness of the sources have only partially been characterized [6]. Studies looking at the influence of variations in eHealth literacy levels [7] and trust in different sources of online health information[2] indicate that the relationship is not linear in all cases, i.e. higher health IQ does not result automatically in higher perceived levels of trustworthiness. For example, in a study by Chen et al. [8], 618 people were recruited to complete a survey which tested their eHealth literacy level, asked them to identify which of 25 sources of health information they used, and how much they trusted each source. The study showed that people with lower eHealth literacy were less likely to trust medical websites (typically higher IQ) and more likely to trust social media, blogs, and celebrity webpages (typically lower IQ) [8].

This might seem a spurious result, were it not for the fact that research has shown that those with high eHealth literacy assess more accurately the credibility and relevance of online health information, whereas those with low eHealth literacy often struggle to locate and understand eHealth information [9]. This difficulty lowers their self-efficacy [10], distorts their perception of source credibility, and impacts negatively perceived trustworthiness [9], ultimately creating a need for individuals with low eHealth literacy to find an alternative means of determining trustworthiness in online sources of information. One such alternative is social endorsement. Visible social endorsement, e.g. 'likes,' [11] enables those with low eHealth literacy to determine trust based on the bandwagon heuristic and assume that, if the source has already been deemed valid by others, then it is safe for them to trust it too [10,12]. Traditional medical websites afford those with low eHealth literacy no such alternative means of determining credibility and trust.

---

[2] eHealth literacy is defined by Norman and Skinner 7.    Norman CD, Skinner HA. eHEALS: The eHealth Literacy Scale. *J Med Internet Res* 2006; **8**(4): e27. as the ability to seek, find, understand, and appraise online health information and apply the knowledge gained to addressing or solving a health problem. This is why we use the term eHealth literacy even though this article focuses exclusively on the IQ of online health information – not broader sources of eHealth information such as electronic health records or wearable devices.

This suggests that those who are more vulnerable to the real-world effects of both disinformation and misinformation (e.g. declining to vaccinate their children) are more likely to rely on online sources with lower health IQ, which are more prone to spread such inflammatory and inaccurate information. Personalisation of online search results may favour this phenomenon and lead to a vicious cycle where the more one searches and reads dis- and mis-information about vaccines the less one finds and reads scientific information on the same topic.

At the same time, there are increasing concerns about the privacy risks associated with Internet search engines storing potentially sensitive and private health information contained within users search histories, combining it with additional information collected for tracking purposes, and using these data for commercial or other purposes [13,14]. This creates a public push back against the idea that search engines or public health providers should interfere in the results people see when searching for health information online. This makes it hard to address concerns about health disinformation/misinformation in a way that is at least socially acceptable if not ideally preferable [15]. The UK's National Health Service (NHS) discovered this when its announcement of Amazon Alexa responding with guaranteed high-IQ content from NHS.UK to user voice queries such as "Alexa, how do I treat a migraine," resulted in a public outcry over privacy infringement [16]. This raises the question whether, in the context of online vaccination information, it is possible at all to balance concerns about user privacy and IQ.

The following pages seek to lay the foundations for answering this question by reporting on the methodology for, and results of, a study concerning the following research question: "What is the current relationship between search engines' approach to privacy and the scientific quality of the information they return?".

**Methods**

In seeking to answer the above research question, this study builds on previous work which:

(a) focuses on the results returned by a search of "vaccines autism" in different countries and languages, and found a great variability in the composition of the webpages returned and their ranking [17]; and

(b) suggests that whether a webpage promotes a medical treatment based on the practices of evidence-based medicine (EBM; e.g. based on the reporting by Cochrane reviews) can act as a proxy for health IQ [18].

The phrase "vaccines autism" was searched using the main search engines (Google, Bing and Yahoo) and alternative providers which focus on respecting privacy and personal data (Duckduckgo, Mojeek, Qwant and Ecosia[3]) in four languages/countries (English-UK, Italian, Spanish and French) and the international, US-English version of Google (Google.com). We also used some country- or language-specific search engines. Each webpage returned was classified as vaccine-positive, -negative or –neutral as described in the Table 1

**Web search and content analysis**

Searches in English were done from Falmer, Sussex, United Kingdom; in Italian from Urbino, Italy; in French from Bruxelles, Belgium; in Spanish from Barcelona, Spain. Each search was done using a logged-out Chrome browser cleared of cookies and previous search history so that the only identification was the IP address and its geolocalization. Additionally, when available, the local version of each search engine was used (e.g. Google.co.uk and Google.it). For searching Google.com, automatic redirection to google.co.uk was avoided by using the URL Google.com/ncr (no-country-redirect).

The first 30 URL results from each search engine result page (SERP), excluding those marked as advertisements, and were transferred to a spreadsheet. When a duplicate URL was encountered, it was excluded. Websites were then visited and the content of each page was coded as vaccine- positive, negative or neutral, depending on the stance taken on the connection between vaccines and autism.

Webpages recommending vaccination and/or negating the link with autism were coded as "vaccine-positive". Those promoting vaccine hesitancy, cautioning about the risk of autism or openly anti-vaccine, were coded as "vaccine-negative". Additionally, webpages that claimed further studies needed to be conducted to clarify the link between vaccines and autism were also coded as "vaccine-negative", as previous research by {Browne, 2015 #421} has shown that users perceive this as confirmation of the fact that vaccine safety has not been proven. Webpages simply reporting the history of the anti-vaccine movement or a related legal debate were coded as "vaccine-neutral". Examples of positive, negative and neutral webpages are provided in Table 1.

---

[3] Duckduckgo.com, states that "our data shouldn't be for sale. Mojeek.com, promise "independent and unbiased search results". Qwant also promises "no discrimination and no bias" and states that "algorithms are applied equally everywhere and for every user, without trying to put websites forward or to hide others based on commercial, political or moral interests" (about.qwant.com). Ecosia has an option to turn off personalization of results and, additionally, responds to ethical concerns related to the environment, and uses its profits to plant trees (https://blog.ecosia.org/ecosia-vs-google-free-alternative-search-engine-taxes-environment-privacy/).

Coding was completed by two raters for each language and inter-rater agreement was calculated with GraphPad, which uses equations 18.16 to 18.20 from Fleiss [19]. On a sample of 59 webpages in English, agreement was 85%, with a Kappa of 0.669 (standard error, 0.077) and a 95% confidence interval from 0.518 to 0.820, a strength of agreement is considered to be 'good' [19]. In Italian, agreement was 90%, with a Kappa of 0.818 (standard error, 0.067) and a 95% confidence interval from 0.687 to 0.950, a strength of agreement is considered to be 'very good'. In Spanish, agreement was 83%, with a Kappa of 0.609 (standard error, 0.098) and a 95% confidence interval from 0.418 to 0.801, a strength of agreement is considered to be 'very good'. In French, agreement was 89% with a Kappa of 0.746 (standard error, 0.091) and a 95% confidence interval from 0.568 to 0.924.

When frequencies of vaccine-negative webpages was compared across different search engines, we used a two-tailed Fisher's test corrected for multiple comparison using the method of Benjamini, Krieger and Yekutieli and a false discovery rate (FDR) of 5%.

**Results**

Figure 1 shows the ranking of positive (green), neutral (yellow), and negative (red) websites returned by the different search engines in English, French, Italian and Spanish.

A comparison of the percentage of vaccine-negative webpages in different SERPs (Table 2) shows that Google is consistently returning less misinformation, although some of its local version show more vaccine-negative webpages than the English-US version (Google.com). Other search engines return some vaccine-negative webpages with some, like Mojeek in English-UK and Arianna and Virgilio in Italian are the most likely to rank higher webpages with misinformation. In English, the frequency of vaccine-negative webpages in Yahoo, Bing and Mojeek was significantly higher than in Google.com. In Italian, all SERPs had a significantly higher proportion of vaccine-negative webpages compared with Google.com.

In French, all search engines returned a significantly higher number of vaccine-negative webpages than Google.com or the local version Google.fr. There were no significant differences for search engines in Spanish.

The SERPs of all search engines providers contained a higher proportion of vaccine-negative results than those obtained from Google.com. Even the localized versions of Google (Google.co.uk, Google.it and Google.es) returned more negative pages than the US/international English Google.com. The two Italian-only search engines (Virgilio and Arianna) returned the highest number of negative pages in Italian.

In several instances, the same webpages were returned by different search engines, with large overlaps between these results as shown in Figure 2.

**Discussion**

The aim of this research was to investigate whether search engines focused on privacy preservation ranked webpages with higher and/or lower levels of IQ (interpreted as EBM-based content i.e. vaccine-positive rather than vaccine-negative) differently to the primary commercial search engines. In turn, this answered the research question "What is the current relationship between search engines' approach to privacy and the scientific quality of the information they return?".

The results indicate that currently privacy-enhancing search engines often give more visibility to webpages promoting vaccine hesitancy or with a clear anti-vaccine position than Google. This is in agreement with the findings from a recent study by the Economist which analyzed 175,000 news links returned by Google to demonstrate that the search engine's algorithm favors trustworthy publications [20]. Reputation and trustworthiness are key factors included in Google's ranking algorithm. In 2019, Google published search quality evaluation guidelines,[4] which define webpages containing information that may affect the users' health or financial stability, as "your money your life" (YMYL) pages. These guidelines reveal that when rating YMYL webpages, Google looks at the three criteria of Expertise-Authority-Trustworthiness (E-A-T) and states

> "High E-A-T information pages on scientific topics should be produced by people or organizations with appropriate scientific expertise and represent well-established scientific consensus on issues where such consensus exists".

Thus, in the case of Google, the bandwagon heuristic mechanism of assessing medical information credibility is working in favour of promoting IQ. This is perhaps because those responsible for driving up the 'reputation' of specific websites by, for example, linking to them, have higher levels of eHealth literacy and therefore act as pseudo-gatekeepers protecting those with lower eHealth literacy from poor IQ results by ensuring that websites providing inaccurate online health information have a low ranking in the search results.

In the case of privacy-preserving engines this appears not to be working – potentially because they do not track 'reputation' factors, which could be seen as proxies for IQ and credibility, e.g. clicks and bounce-rate, and so have no gatekeepers (pseudo or otherwise) working to limit the circulation of misinformation. To our knowledge, the algorithms used by these 'alternative' search engines are not public - Qwant states that they use their own algorithms[5] - but despite this, we found a large overlap between the SERP of Qwant and those of Bing and Ecosia. Likewise, Ecosia and Swisscows showed a similar overlap (70%) with Bing. This means that it is not possible to check what factors determine the outcome of the decision making processes of the search engines algorithms, and hence identify those elements that contribute to circulate misinformation.

---

[4] (https://static.googleusercontent.com/media/guidelines.raterhub.com/en//searchqualityevaluatorguidelines.pdf)
[5] https://medium.com/qwant-blog/web-indexation-where-does-qwants-independence-stand-8eab4f7856f8
(Archived at: https://web.archive.org/web/20190627090553/https://medium.com/qwant-blog/web-indexation-where-does-qwants-independence-stand-8eab4f7856f8)

Even without being able to check the exact mechanisms for the overall poorer results, the results suggest that, currently, decisions made by search engines to prioritize privacy-preservation may have a negative impact on the IQ of results returned to users in health contexts. The pledge to provide independent and unbiased results or not to promote or hide websites based on political or moral interests can be seen as ethically ambiguous, in view of the potential consequences of pointing to scientifically unsound health information.

Medical ethics requires that patients give informed consent before treatment and must, therefore, be informed accurately of the risks and benefits associated with treatment, something that is not possible if the search engine provider used by an individual is returning results with low IQ. From this consequentialist perspective, it is inherently unethical choosing not to interfere with a search engine's ranking algorithm to ensure 'manually' that results of higher IQ are prioritized, whilst those of lower IQ are suppressed. This builds on arguments already made in the wider literature about algorithm ethics [21,22]. For example, the Association for Computing Machinery (ACM) code of ethics and professional conduct includes, as a general principle, "avoid harm" along with that of "respect privacy" [23], which aligns with the Hippocratic oath. Specifically, the implication that the current design of privacy-preserving search engine algorithms underestimates the need for evaluation of IQ fits into the wider discussion about algorithmic design and how to ensure design decisions are made to protect and incorporate key values such as IQ. It is necessary, but insufficient, to design search engine algorithms that index purely on 'relevance', they must also be designed to index on quality. The challenge lies in the ability to do this in a way that balances the need to accept different perspectives (particularly those that are rooted in different cultural, religious or social ideals) whilst also filtering for IQ. Supporters of such arguments, including the authors, note that this requirement necessitates making the workings of search engine algorithms more transparent [24] to ensure their ethical compliance.

Providing information on vaccines that is based on misinformation or disinformation (including studies whose data or conclusions have been shown to be wrong) is a deceptive practice, which goes against the basic tenets of medical and business ethics [25]. This is in line with those who argue that the promotion of 'alternative medicine' is unethical because it lacks evidence and transparency of clinical efficacy and should be considered 'false advertising' [26]. Online service providers have a moral responsibilities to ensure that users access health information that is scientifically validate [27]. Misinformation and disinformation concerning healthcare circulating on the internet can have severe consequences and lead to widespread harm. Consider the example of South African president Mbeki, who delayed introducing anti-retroviral drugs in favor of alternative medicine based on information obtained from HIV-denialist internet websites, a decision estimated to have resulted in over 300,000 deaths [28]. More recently, a cancer patient died in China after following an alternative, non-approved therapy they found using a search engine [29]. In response Chinese authorities issued new rules that require search engines to provide "objective, fair and authoritative results".

Moral responsibility follows on the level of harm that misinformation and disinformation on healthcare may cause. It could be argued that there is a greater onus on Google – and other commercial search engine providers – than on alternative search engines to take these considerations into account when designing or interfering with algorithms for the purposes of promoting ethical compliance, given their larger market share. Google currently has over 90% of the worldwide market share[6] and therefore has the potential to indirectly 'cause harm', through the potential promotion of low IQ webpages on vaccinations, to a great many more people than any of the alternative providers.

However, studies have shown that those who hold anti-authoritarian views, openness to (potentially) controversial opinions, and an interest in alternative medicine are, in some cases, already more likely to hold vaccine-hesitant beliefs [30]. These are also the individuals who are most likely to use alternative search providers given their sensitivity to privacy and tracking concerns. Therefore, while the alternative providers might reach a proportionally smaller audience – they are reaching an audience that is already more receptive to anti-vaccine information [31], and therefore more vulnerable to its effects. In other words, unless the alternative providers take steps to rank vaccine-related search results according to IQ, they may cause greater harm to those they do reach, meaning that the net negative impact is still greater even though the number of individuals they reach is smaller.

---

[6] https://gs.statcounter.com/search-engine-market-share

**Conclusion**

Our analysis shows that while it may well be technically possible to design a search engine that manages to balance privacy-preservation with the promotion of high IQ material, this is currently not the case. The current relationship between privacy-preserving design features of search engines and the IQ of the results they return is inverse (although not proportionally). In instances where this can have a negative impact on public health, as in the example we have provided of the promotion of anti-vaccine misinformation, not intervening to alter the design of the algorithm - even if this means sacrificing some degree of user privacy - can lead to severe harm for a large population of user and is, therefore, unethical.

Designing a search engine that is privacy savvy and avoids issues with filter bubbles that can result from user-tracking may be a good thing, like designing a car with an engine that does not pollute and is inexpensive to run, and designers should seek to balance the different aspects of search engine design highlighted in Figure 3. However, if the brakes in an environmentally-designed car do not work, the car is unsafe and this negates the positive ethical decisions made by the designers. In a car this is a highly unlikely design outcome, as a car has to undergo several rounds of testing by regulatory agencies before being allowed on the market. This is not yet the case for search engines, which are only regulated from the perspective of data protection – which is primarily interpreted as data security rather than data privacy. Our study suggests that this is necessary but insufficient, and instead mechanisms should be developed to test search engines from the perspective of IQ, (particularly for YMYL webpages) before they can be deemed trustworthy providers of public health information.

**Table 1. Example of classification of webpages**

| **Positive stance on vaccines** |
|---|
| https://www.historyofvaccines.org/content/articles/do-vaccines-cause-autism |
| https://www.nhs.uk/news/medication/no-link-between-mmr-and-autism-major-study-finds/ |
| https://www.nhs.uk/conditions/vaccinations/mmr-vaccine/ |
| https://www.autism.org.uk/get-involved/media-centre/position-statements/mmr-vaccine.aspx |
| https://kidshealth.org/en/parents/autism-studies.html |
| https://vaccine-safety-training.org/mmr-vaccine-increases.html |
| https://www.nejm.org/doi/full/10.1056/NEJMoa021134 |
| https://www.ncbi.nlm.nih.gov/pubmed/30986133 |
| https://www.ncbi.nlm.nih.gov/pubmed/15366972 |
| **Negative stance on vaccines** |
| https://www.thelancet.com/journals/lancet/article/PIIS0140673605756968/fulltext |
| www.whale.to/vaccine/vaccine_autism_proven.html |
| www.vaccineriskawareness.com/Infant-Vaccines-Produce-Autism-Symptoms-In-Primates |
| https://leftbrainrightbrain.co.uk/2014/07/17/more-of-that-vaccineautism-research-that-doesnt-exist/ |
| edition.cnn.com/2011/HEALTH/01/05/autism.vaccines/index.html |
| https://www.coasttocoastam.com/show/2018/02/21 |
| nocompulsoryvaccination.com/2014/08/22/vaccine-autism-cover-up/ |
| https://www.newswars.com/vaccine-autism-questioned-by-doctor-congressman-elect/ |
| vaxtruth.org/2011/08/vaccines-do-not-cause-autism/ |
| **Neutral stance on vaccines** |
| https://www.physicsforums.com/threads/vaccine-and-autism-link.880852/ |
| www.discovermagazine.com/2009/jun/06-why-does-vaccine-autism-controversy-live-on |
| https://www.ecso.org/news/autism-charity-founder-anti-vaccination-campaigner/ |
| https://www.theguardian.com/society/2019/jun/01/professor-who-links-vaccines-to-autism-funded-through-university-portal |
| https://www.independent.co.uk/news/world/americas/trump-vaccines-autism-links-anti-vaxxer-us-president-false-vaccine-a8331836.html |
| https://www.ncbi.nlm.nih.gov/pmc/articles/PMC2376879/ |

**Table 2. Percentage of vaccine-negative webpages in different SERPs**

|              | English | Italian | Spanish | French |
|--------------|---------|---------|---------|--------|
| **Google.com**   | 0       |         |         |        |
| **Local Google** | 7       | 10      | 10      | 0      |
| **Yahoo**        | 23*     | 30*     | 13      | 23*    |
| **Bing**         | 23*     | 27*     | 17      | 20*    |
| **Duckduckgo**   | 17      | 30*     | 17      | 23*    |
| **Ecosia**       | 10      | -       | 20      | 23*    |
| **Qwant**        | 13      | 33*     | 17      | 23*    |
| **Mojeek**       | 53*#    | -       | -       | -      |
| **Swisscows**    | 17      | -       | -       | -      |
| **Arianna**      | -       | 37*     | -       | -      |
| **Virgilio**     | -       | 37*     | -       | -      |

Total of 30 webpages for each SERP. Cells are colour coded: median (20%), yellow; maximum (53%), red; minimum (0%), green. *Significantly different from Google.com by two-tailed Fisher's test corrected for multiple comparison using the method of Benjamini, Krieger and Yekutieli at a false discovery rate of 5%. #significantly different from the localized version of Google.

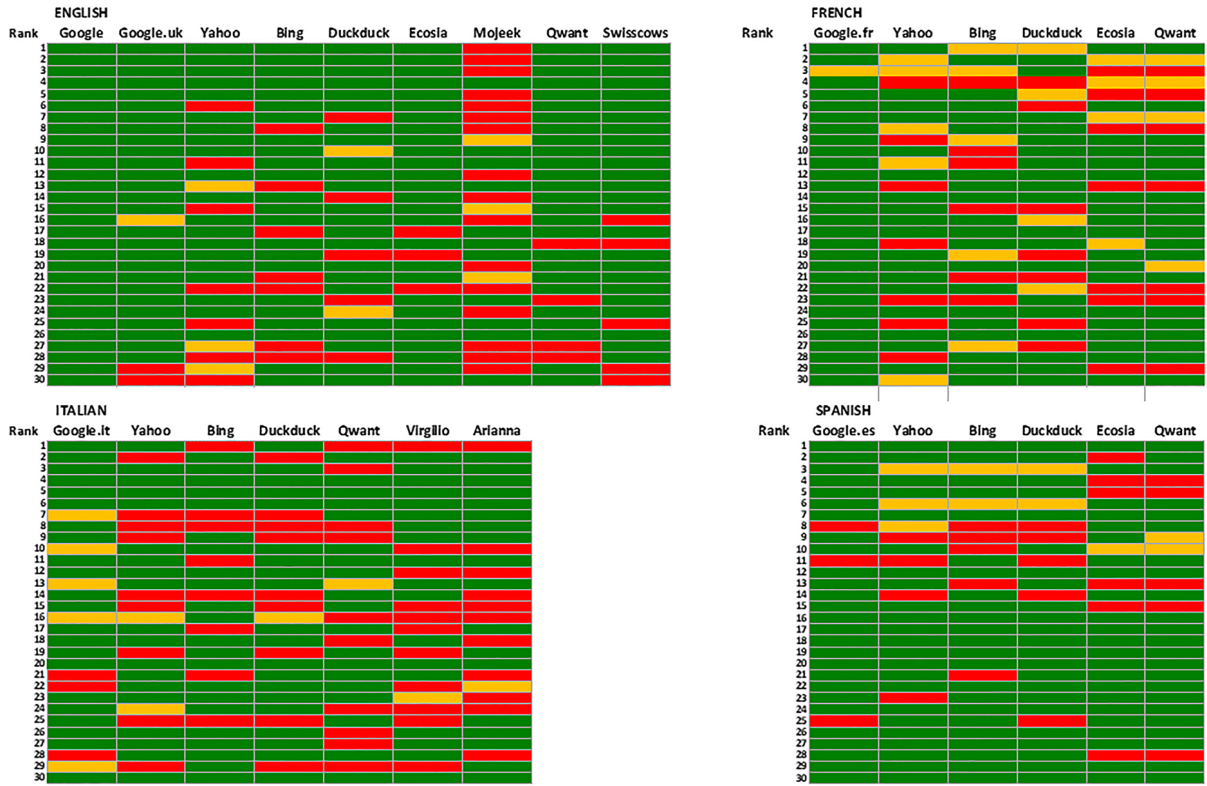

**Figure 1. Stance on vaccines in webpages returned by different search engines in four languages**. The top 30 webpages returned in the SERPs are shown. Green, vaccine-positive, yellow, vaccine-neutral, red, vaccine-negative.

## ENGLISH

| | Google.com | GoogleUK | Yahoo | Bing | Duckduck | Ecosia | Mojeek | Qwant | Swisscows |
|---|---|---|---|---|---|---|---|---|---|
| Google.com | 30 | 8 | 6 | 7 | 7 | 5 | 1 | 6 | 5 |
| GoogleUK | | 30 | 6 | 4 | 4 | 4 | 0 | 4 | 4 |
| Yahoo | | | 30 | 11 | 13 | 9 | 2 | 9 | 10 |
| Bing | | | | 30 | 11 | 21 | 1 | 21 | 23 |
| Duckduck | | | | | 30 | 13 | 0 | 13 | 14 |
| Ecosia | | | | | | 30 | 1 | 28 | 26 |
| Mojeek | | | | | | | 30 | 1 | 1 |
| Qwant | | | | | | | | 30 | 26 |
| Swisscows | | | | | | | | | 30 |
| Unique | 12 | 18 | 10 | 3 | 7 | 1 | 28 | 1 | 1 |

## FRENCH

| | Google.fr | Yahoo | Bing | Duckduckgo | Ecosia | Qwant |
|---|---|---|---|---|---|---|
| Google.es | 30 | 4 | 5 | 5 | 5 | 1 |
| Yahoo | | 30 | 25 | 12 | 10 | 0 |
| Bing | | | 30 | 11 | 24 | 0 |
| Duckduck | | | | 30 | 12 | 0 |
| Ecosia | | | | | 30 | 1 |
| Qwant | | | | | | 30 |
| Unique | 22 | 1 | 4 | 16 | 1 | 29 |

## ITALIAN

| | Google.it | Yahoo | Bing | Duckduck | Qwant | Virgilio | Arianna |
|---|---|---|---|---|---|---|---|
| Google.it | 30 | 5 | 4 | 5 | 3 | 2 | 1 |
| Yahoo | | 30 | 7 | 20 | 27 | 9 | 9 |
| Bing | | | 30 | 11 | 4 | 13 | 13 |
| Duckduck | | | | 30 | 17 | 13 | 13 |
| Qwant | | | | | 30 | 10 | 12 |
| Virgilio | | | | | | 30 | 28 |
| Arianna | | | | | | | 30 |
| Unique | 23 | 0 | 12 | 5 | 0 | 1 | 0 |

## SPANISH

| | Google.es | Yahoo | Bing | Duckduck | Ecosia | Qwant |
|---|---|---|---|---|---|---|
| Google.es | 30 | 10 | 8 | 10 | 2 | 2 |
| Yahoo | | 30 | 26 | 27 | 5 | 5 |
| Bing | | | 30 | 26 | 4 | 4 |
| Duckduck | | | | 30 | 5 | 5 |
| Ecosia | | | | | 30 | 30 |
| Qwant | | | | | | 30 |
| Unique | 20 | 1 | 2 | 1 | 0 | 0 |

**Figure 2. Overlaps among different SERPs**

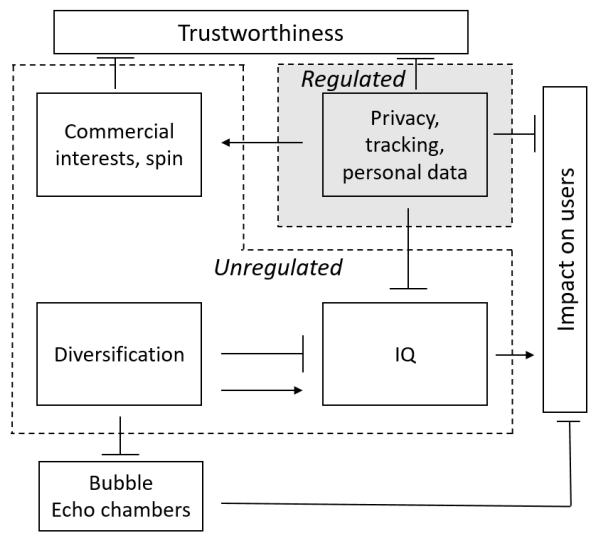

**Figure 3. A summary of the different aspects of search engines ranking and their relationship.**